\documentclass[11pt]{article}
\usepackage{amsmath}
\usepackage{amssymb}
\usepackage{graphicx}
\usepackage{caption2}
\usepackage{amsfonts}
\usepackage{cite}

\newcommand{\be}{\begin{equation}}
\newcommand{\ee}{\end{equation}}
\newcommand{\bea}{\begin{eqnarray}}
\newcommand{\eea}{\end{eqnarray}}
\newcommand{\bef}{\begin{figure}}
\newcommand{\ef}{\end{figure}}
\newcommand{\bt}{\begin{tabular}}
\newcommand{\et}{\end{tabular}}
\newcommand{\bno}{\begin{enumerate}}
\newcommand{\eno}{\end{enumerate}}

\def\3{\ss}

\catcode`\"=\active
\def"{\accent'177}
\begin{document}

\onecolumn

\begin{center} 

{\bf\Large On the mathematically reliable long-term simulation of chaotic solutions  
 of Lorenz equation  in the interval [0,10000]}

Shijun Liao $^{a,b,c}$  \footnote{Corresponding author, email address: sjliao@sjtu.edu.cn}   

$^a$ State Key Laboratory of Ocean Engineering, Shanghai 200240, China\\

 \vspace{0.25cm}

$^b$ School of Naval Architecture, Ocean and Civil Engineering\\
Shanghai Jiao Tong University, Shanghai 200240,  China\\

 \vspace{0.25cm}
 
$^c$ Nonlinear Analysis and Applied Mathematics Research Group (NAAM) \\  
King Abdulaziz University (KAU), Jeddah, Saudi Arabia

\vspace{0.3cm}

Pengfei  Wang  

State Key Laboratory of Atmospheric Sciences and Geophysical Fluid Dynamics\\
Institute of Atmospheric Physics, Chinese Academy of Sciences, Beijing, China

\end{center}

{\em 
Using 1200 CPUs  of  the National Supercomputer TH-A1 and a parallel integral algorithm based on  the  3500th-order Taylor expansion  and  the 4180-digit  multiple precision data,  we have done a reliable simulation of  chaotic solution of Lorenz equation in a rather long interval $0 \leq t \leq  10000$ LTU (Lorenz time unit).  Such a kind of  mathematically reliable chaotic simulation has never been reported.      It provides us a numerical benchmark for  mathematically  reliable long-term prediction of chaos.        Besides,  it also proposes a safe method for  mathematically  reliable  simulations of chaos in a finite but long enough interval.   In addition,  our very fine simulations suggest that such a kind of  mathematically reliable long-term prediction of chaotic solution  might  have  no   physical  meanings,  because  the   inherent  physical  micro-level  uncertainty due to thermal fluctuation might quickly transfer into macroscopic uncertainty so that trajectories for a long enough time would be essentially uncertain in physics.  }         
 

\section{Introduction}

Lorenz \cite{Lorenz1963} found that the nonlinear dynamic system (called today Lorenz equations) 
\begin{eqnarray}
\dot{x} &=& \sigma \left( y - x \right), \\
\dot{y} &=& R \; x - y - x \; z, \\
\dot{z} & = & x \; y - b \; z
\end{eqnarray}
has non-periodic solution in cases such as  $ b = 8/3$, $\sigma=10$ and $R > 24.74$,  and especially,  numerical simulations  are rather sensitive to the initial conditions, i.e. a very slight variation in the initial condition of Lorenz equation might lead to a significant difference of numerical simulation.  Lorenz's work \cite{Lorenz1963}  in 1963 was a milestone.  Today, it is common knowledge  that, due to the so-called sensitive dependence on initial condition (SDIC) or the so-called ``butter-fly effect'', it is impossible to make a long-term prediction of chaotic dynamic systems \cite{Lorenz1963, Lorenz1989, Lorenz2006, Egolf2000, Gaspard}.  Here, the prediction means that one can make {\em reliable},  convergent numerical simulations.  In other words, due to  ``the butter-fly effect'', chaotic systems destroy the possibility of following the true individual trajectories  accurately in an {\em infinite} interval of time.           

Shadowing theories  have proliferated in the literature \cite{Anosov1967}   to locate or prove the existence of true trajectories that stay near the computer-generated  trajectory for a long time.   According to the Shadow Lemma \cite{Anosov1967},  for a uniformly hyperbolic dynamic system, there always exists a true trajectory near any computer-generated trajectories, as long as the truncation and round-off errors are small enough.  Unfortunately, hardly a nonlinear dynamic system is  uniformly hyperbolic in most cases so that the  Shadow Lemma \cite{Anosov1967}  seldom  woks  in practice.   Besides,  it is found that  for some chaotic dynamic systems,   computer-generated trajectories can be shadowed only for a short time \cite{Yorke1994},  and in addition  it is ``virtually impossible to obtain a long trajectory that is even approximately correct'' \cite{Yorke1997}.  Furthermore,  for some chaotic systems,  ``there is no fundamental reason for computer-simulated long-time statistics to be even approximately correct'' \cite{Sauer2002}.   As illustrated by Yuan and York \cite{Yorke2000} using a model, a numerical artifact persists for an arbitrarily high numerical precision.   These numerical artifacts ``expose an exigent demand of safe numerical simulations''  \cite{Shi2008}.   

Note that floating-point calculations to approximate solutions of dynamic systems contain the inherent numerical noises, say, the truncation and round-off errors.  Lorenz \cite{Lorenz1989} investigated the sensitivity of numerical simulations to the time step and found the so-called ``computational chaos''.     Li et al. \cite{Li2000} investigated the influence of the time step $\Delta t$ on the numerical simulation of Lorenz equation in details.   They used  29 standard numerical methods at orders up to 10  in single and double precision.    In case  of   $b= 8/3$, $\sigma=10$ and $R < 24.06$ (without chaos), they found  \cite{Li2000} that the final status of computer-generated trajectories starting from (5,5,10) are rather sensitive to the time step $\Delta t$ varying from $10^{-6}$ to $10^{-1}$:  it seems to alterlate  randomly  between the two unstable fixed points 
\[  C\left(\sqrt{b (R-1)},  \sqrt{b (R-1)}, R-1\right) ,  C'\left(-\sqrt{b (R-1)},  -\sqrt{b (R-1)}, R-1\right)   \]  
with the same  probability. So, in some cases even without chaos,  it is also impossible to accurately predict the final status of Lorenz equation using the 29 traditional numerical methods  at orders up to 10  in single or double precision \cite{Li2000}.   This numerical phenomenon  \cite{Li2000,Li2001}  revealed  the significant influence of truncation and round-off errors on nonlinear dynamic systems.   This kind of  sensitivity to time step was further studied and confirmed by Lorenz \cite{Lorenz2006} in 2006,  Teixeira et al. \cite{Teixeira2007} in 2007 and Liao \cite{Liao2009} in 2009.  It should be emphasized that the truncation and round-off errors are ineluctable.   Therefore,   numerical simulations of chaotic dynamic systems are sensitive not only to initial condition but also to numerical algorithms and numerical precision  at each time step.   

How long is a numerical simulation reliable?  This is an important question, as pointed out by Sauer et al. \cite{Yorke1997} who proposed the concepts of ``shadowing time''  to answer the question.  As illustrated by Li et al. \cite{Li2000},  different chaotic numerical simulations of Lorenz equation gained by means of the  29 standard  numerical methods  agree   with  each  other  only  in  the  interval  $[0,T_c]$,  where $T_c \approx 16.857$ for single precision data and $T_c \approx 35.412$ for double precision data, respectively.    Li et al. \cite{Li2000, Li2001} proposed the concept ``maximally effective computation time'' (MECT), beyond which numerical simulations obtained by different time steps have significant difference and thus should be unrelated to the true solution.   Teixeira et al. \cite{Teixeira2007} proposed the concept of ``critical time of decoupling'',  $T_c$, defined as the first point in time after which the state vector norm error exceeds a certain threshold.   They  \cite{Teixeira2007}  observed that the critical time of decoupling $T_c$ is directly proportional to  $\ln (\Delta t)$, where $\Delta t$ denotes the time step.   The concepts of ``shadowing time'',  ``maximally effective computation time'' and  ``critical time of decoupling'', although defined in different ways,   reveal the same fact:   it is only possible to give the reliable numerical simulations of chaotic systems in a {\em finite} interval $[0,T_c]$, beyond which numerical artifacts might occur.   Then, the key point is to ensure that such a kind of interval is long {\em enough}, i.e. $T_c$ is large enough.         

As pointed out by Wang et al.  \cite{Wang2011},  in  order to gain a reliable chaotic solution of Lorenz equation in the interval [0,1000] by means of the traditional 4th-order Runge-Kutta method,  one had to use a rather small time-step $\Delta t = 10 ^{-170}$ and very accurate data in 10000-digit precision, but the required  CPU time is about $3.1\times 10^{160}$ years,  which is even longer than the existence of our universe up to now.   Thus,  in  order  to  gain a reliable chaotic simulation in a long enough interval  within  acceptable CPU times,  both of the truncation and round-off errors must be small enough.  Therefore,  not only  the precision of data but also the order of  numerical algorithms must be high enough.     
                            
Based on the Taylor series method \cite{Lorenz2006, Corliss1982, Barrio2005}  at high {\em enough} order and  data  in  high {\em enough} precision,  the so-called  ``Clean Numerical Simulation'' (CNS)  is developed by Liao \cite{ Liao2009, Liao2013}   to  gain mathematically  {\em reliable}  simulations of chaotic dynamic systems in a {\em finite} but long {\em enough} interval of time.   The Taylor series method \cite{Lorenz2006, Corliss1982, Barrio2005} is one of the oldest methods, which traces  back to Newton, Euler, Liouville and Cauchy.  It has an advantage that its formula at an arbitrarily high order can be easily expressed in the same form.  So, from the viewpoint of numerical simulations,  it is rather easy to use the Taylor series method at a very high order so as to deduce the truncation error to a required level.  At the very beginning of the CNS \cite{Liao2009}, the computer algebra system Mathematica was used to decrease the round-off error to a required level,  because Mathematica can express all numerical data in arbitrarily high precision.  Let $M$ denote the order of the Taylor series method, and $N_s$ the significant digit number of all numerical data, respectively.   Unlike other numerical approaches, in the frame of the CNS,  the  significant  digit  number $N_s$ increases linearly with $M$ (the order of Taylor expansion),  for example $N_s = 2M$ as illustrated in \cite{Liao2009}, so that both of the truncation and round-off errors decrease samutaneously.  In this way, both of the truncation and round-off errors can be reduced to a required level by means of high enough order of Taylor series expansion and high enough precision of data,  so that the numerical simulations of chaos are  mathematically reliable in a time interval $[0, T_c]$, where $T_c$ is called {\em the critical prediction time}.   

In the frame of the CNS,  the critical prediction time $T_c$ is dependent upon the  order $M$ of the Taylor expansion and the significant digit number $N_s$.   Let $s( M, N_s)$ denote a numerical simulation given by the $M$th-order Taylor series method in $N_s$ digit precision.   Let $T_d(M,N_s; M',N'_s) $ denote the time of decoupling, determined by comparing $s( M, N_s)$ with a more accurate simulation $s'( M', N'_s)$, where $M' > M$ and $N'_s \geq N_s$.  Then, the critical prediction time $T_c$ of the numerical simulation $s(M,N_s)$ is defined by
\begin{equation}
T_c(s) = \min_{M'>M, N'_s > N_s} T_d (M,N_s; M',N'_s) 
\end{equation}    
for {\em any} $M'>M$ and $N'_s \geq  N_s$.  

Another key point of the CNS is that an explicit estimation of the critical prediction time $T_c$ versus the order $M$ of the Taylor series expansion is given, as illustrated by  Liao \cite{Liao2009} who gave the estimation $T_c \approx 3M$ for Lorenz equation (using data in $2M$-digit precision, i.e. $N_s = 2M$).   Then, given a finite but long {enough} time interval $[0,T_c]$ of Lorenz equation, one might obtain  mathematically reliable numeral simulations of chaotic solution in  the interval $[0,T_c]$ by means of the CNS with the order  $M > T_c/3$ of Taylor expansion and data in $(2M)$-digit precision, which however should be verified by means of a higher order of Taylor series method.      

In this paper, we gave a mathematically reliable chaotic numerical simulation of Lorenz equation in the interval [0,10000] by means of the CNS using parallel computation in the National Supercomputer TH-A1.  To the best of our knowledge, mathematically reliable chaotic simulations in such a long interval have never been reported.  So, it provides us a numerical benchmark for reliable long-term simulation of chaos.   Thereafter, we point out that, due to the thermal fluctuation,  the physical micro-level uncertainty of initial conditions is inherent.  This objective, inherent,  physical uncertainty of initial condition is indeed rather small (at the level of $10^{-19}$), but is much larger than the required 4000-digit precision for a reliable prediction of chaotic solution in such a long interval $[0,10000]$.  This suggests that  the  chaotic trajectories of Lorenz equation for a long time might be essentially uncertain in physics.                

\section{Reliable long-term numerical simulation of chaos}

Liao \cite{Liao2009}  proposed the CNS to gain reliable chaotic solutions of Lorenz equation in the case of $R = 28,  b = 8/3$ and $\sigma=10$.  Using time step $\Delta t = 0.01$ and enforcing the significant digit number $N_s=2M$, where $M$ denotes the order of Taylor series method, Liao \cite{Liao2009}  found an estimation formula $T_c \approx 3M$ for the criticial prediction time $T_c$.  According to this formula, Liao  \cite{Liao2009} gained a reliable chaotic solution of Lorenz equation in the interval $0 \leq t \leq 1000$ Lorenz time units (LTU) by means of the CNS using the 400th-order Taylor series method and the data in 800-digit precision.   It took about one month \cite{Liao2009}, mainly due to the use of the computer algebra system Mathematica  without parallel computation.  The computational efficiency of the CNS was   improved  greatly (several hundreds times faster) by Wang et al \cite{Wang2011}  who employed a parallel algorithm and the multiple precision (MP) library of C.  Their result  (based on  the 1000th-order Taylor expansion and 2100-digit MP precision) confirms the correction of Liao's simulation \cite{Liao2009} in the interval $[0,1000]$.   Using the CNS with the 1000th-order Taylor expansion and the 2100 digit multiple precision,  Wang \cite{Wang2011}  obtained  a reliable chaotic solution in the interval [0,2500] within only 30 hours, which is validated using a more accurate simulation given by the 1200th-order Taylor expansion and the 2100 digit multiple precision.       

\begin{table}[t]
\caption{Reliable results of Lorenz equation in the case of $\sigma = 10,  R = 28,  b = -8/3$ and $x(0) = -15.8,  y(0) = -17.48,  z(0) = 35.64$   by means of the parallel algorithm of the CNS with the 3500th-order Taylor expansion,  the 4180-digit MP  data and $\Delta t = 0.01$.  The used CPU times is 9 days and 5 hours by means of  the 1200 CPUs of National Supercomputer TH-1A.}
\begin{center}
\begin{tabular}{|c|c|c|c|} \hline\hline 
$t$ & $x$ & $y$ & $z$ \\ \hline
 500 	& -5.3050   & -9.4260 	& 12.3022\\
1000 & 13.8820 & 19.9183	& 26.9019\\
1500& -10.1398	& -7.6264	& 31.8584\\
2000& -6.8739	& -1.4848	& 31.3495\\
2500& 2.7592	& 0.4763		& 24.6411\\
3000& 1.6933	& 3.6003		& 21.4109\\
3500& 0.7357	& -2.1187	&24.4667\\
4000& -7.6927	& -13.4996	& 14.1994\\
4500& -13.7455	& -8.3158	& 38.8589\\
5000& -6.0844	& -10.8137	& 12.7391\\
5500& 4.7719	& 8.8154		& 10.4386\\
6000& 0.2167	& 2.1043		& 22.1246\\
6500& 4.6758	& 5.6919		& 20.4906\\
7000& -11.3949	& -16.5754	& 23.6813	\\
7500& 0.1858	& 0.6489		& 16.5550\\
8000& -1.2659	& -2.3363	& 17.4960\\
8500& -3.0412	& 1.5314		& 27.8442\\
9000& 13.4797	& 17.2821	& 29.2382\\
9500& 8.9996	& 3.0374		& 33.8242\\
10000& -15.8173	& -17.3669	& 35.5584\\
\hline\hline
\end{tabular}
\end{center}
\label{Lorenz}
\end{table}% 

Can we obtain a mathematically reliable chaotic simulation of Lorenz equation in a longer interval like [0,10000]?  Obviously, such a  reliable chaotic solution can  provide us a numerical benchmark for  {\em mathematically} reliable long-term prediction of chaos, and thus certainly has  important meanings in theory.   

According to Liao's \cite{Liao2009} estimation  $T_c \approx 3 M$ for Lorenz equation in the case of $N_s=2M$,  the required order of Taylor expansion should be larger than 3333 for a reliable simulation in [0,10000].  We found that the required digit precision might be less than $2M$.  Even so,  it  is  a  huge  challenge to gain such a reliable chaotic simulation of Lorenz equation in [0,10000].    Currently, using the parallel algorithm and 1200 CPUs of the National Supercomputer TH-1A (at Tianjian, China),  we successfully obtained the reliable chaotic solution of Lorenz equation  in the interval $[0,10000]$  by means of the CNS using the 3500th-order Taylor expansion  and the 4180-digit  multiple precision.  The used  CPU time is 220.92 hours (i.e. about 9 days and 5 hours).    Its  reliability  (from the mathematical viewpoint) was confirmed by means of the CNS using  the 3600th-order Taylor  expansion  and the 4515-digit multiple precision.   The reliable chaotic solution is given in Table~\ref{Lorenz}, which can be regarded as a benchmark of long-term prediction of chaos.    It should be emphasized that, to the best of our knowledge,  the mathematically reliable chaotic solution of Lorenz equation in such a long interval of time  has {\em never} been reported, whose time interval  of  the  reliable chaotic solution  is about 500 times longer than that of chaotic simulations given by means of the traditional Runge-Kutta method in double-precision.  Therefore, from the {\em mathematical} viewpoint, it is possible to gain a  reliable, convergent chaotic solution of Lorenz equation in such a long interval $[0,10000]$ within a reasonable CPU time, as long as  the initial condition and all data at each time-step are accurate {\em enough}, and besides the order of the Taylor series  method  is high {\em enough}.  

It was found  \cite{Liao2009}  that the required initial condition of Lorenz equation should be in the accuracy of $10^{-0.4 T_c}$, where $T_c$ is the critical prediction time.  Thus, from {\em mathematical} viewpoint,   when $T_c = 10000$, the required initial condition must be in the accuracy of $10^{-4000}$, i.e. in the 4000-digit precision.       However,  from the {\em physical} viewpoint,  does such an  accurate initial  condition  exist  in nature?         
  
\section{Are mathematically reliable long-term chaotic simulations physically meaningful?}

It is traditionally believed that initial conditions are exact in nature, and  the uncertainty  in  initial  conditions  is  due to the fact that we can not measure at arbitrary precision.   Thus,  due to  the ``butterfly effect'',  such a kind of uncertainty or limited knowledge of initial conditions  destroys  long-term prediction of chaos.  This is the traditional explanation  for the ``butter-fly effect'' and SDIC of chaos.  However, this traditional thought is wrong in physics, since the uncertainty of initial conditions of Lorenz equation is {\em objective} and {\em inherent} in nature, as shown below.     

Note that Lorenz equation is a simplified model for convention and heat transfer of viscous fluid \cite{Lorenz1963, Saltzman1962}, with $x,y$ for velocity and $z$ for temperature, respectively.  It should be emphasized that, thermal fluctuation is a basic consequence of the definition of temperature \cite{Khinchin1949, Landau1985, Gorodetsky2004}: a system at nonzero temperature does {\em not} stay in its equilibrium microscopic state, but instead {\em randomly} samples {\em all} possible states, with probabilities given by the Boltzmann distribution.   Thermodynamic variables, such as pressure and temperature, undergo thermal fluctuations in a similar way.   For example, a system has an equilibrium temperature,  but  its  {\em true} temperature fluctuates to some extent about the equilibrium.  

According to statistical mechanics \cite{Khinchin1949, Landau1985},   the well-known thermodynamic equation \cite{Gorodetsky2004} for the variance of temperature fluctuations $u$ in volume $V$  reads
\begin{equation}
\left< u^2\right> = \frac{k_B T^2}{\rho C_v V},
\end{equation}
where $k_B=1.3806488 \times 10^{-23}$ (J/K) is the Boltzmann constant, $T$ denotes temperature,  $\rho$ the density,  and $C_v$  the specific heat capacity, respectively.  
In the case of the typical air in room conditions in a cube with $V=10^{12}$ (m$^3$), we have  $T= 20$ C = 293.15 K, $\rho = 1204.1 $~g/m$^{3}$,  $C_v = 1.012$ J/(g K),   respectively.   The corresponding  standard deviation  of temperature  reads  $\sqrt{\left< u^2\right>} =  3.1204 \times 10^{-17}$ (K).   Using the room temperature $T=293.15$ K as the characteristic one, we have the dimensionless  standard deviation  of temperature  $\sigma = 1.06444 \times 10^{-19}$.   This value is indeed rather small.  However, it is much larger than $10^{-4000}$, the required precision of the initial condition for a {\em mathematically} reliable prediction of chaotic simulation in the interval [0,10000].   

Without loss of generality,  let us consider  the  initial  condition 
\begin{equation}
x(0) = -15.8,  y(0) = -17.48,  z(0) = 35.64 \label{IC:A}
\end{equation}
of Lorenz equation (in the case of $R=28,  b= 8/3$ and $\sigma=10$) for the equilibrium state.   As mentioned above,  a system at nonzero temperature  may  {\em randomly}  sample  {\em all}  possible  states \cite{Khinchin1949, Landau1985, Gorodetsky2004}.  For simplicity,  let us consider  the following two possible states, corresponding to the two  initial conditions  with a  micro-level  thermal fluctuation of temperature (at $t=0$)
\begin{equation}
x= -15.8,  y = -17.48,  z = 35.64 + 10^{-20}     \label{IC:B}
\end{equation}
or 
\begin{equation}
x = -15.8,  y = -17.48,  z = 35.64 - 10^{-20}.     \label{IC:C}
\end{equation}
From the {\em physical} viewpoints of thermal fluctuation, the above three initial conditions (\ref{IC:A}), (\ref{IC:B}) and (\ref{IC:C}) are the {\em same}, since each of them may sample a state   but  we do {\em not} know which ones truly appears  in practice.   The key point is that, {\em physically} speaking,  due to the thermal fluctuation,  the required initial condition in the accuracy of 4000-digit precision for $T_c = 10000$  (LTU) does  {\em not} exist in nature, since it is  much smaller even than the thermal fluctuation!

The CNS provides us a safe tool to very accurately simulate the propagation of this kind of inherent,  objective,  physical,  micro-level uncertainty of initial conditions, as illustrated by Liao \cite{Liao2009, Liao2013, Liao2012CNSNS, Liao2013CNSNS} and Wang et al \cite{Wang2011}.   Similarly,  by means of the parallel CNS  (8 CPUs) with the 400th-order Taylor expansion and the data in accuracy of 1000-digit multiple precision ($\Delta t =10^{-2}$),  we gain the three reliable chaotic solutions in the interval $0\leq t \leq 1000$, corresponding to the possible initial conditions (\ref{IC:A}), (\ref{IC:B}) and (\ref{IC:C}),  respectively, within one and a half  hour.   Besides,  the  mathematical reliability of these three chaotic simulations  in the interval $t \in [0,1000]$  is confirmed by means of  the 500th-order Taylor expansion and the 1200-digit multiple precision data.  In this way,  these three chaotic simulations  are guaranteed to be {\em mathematically} reliable in the interval $0\leq t \leq 1000$.   In other words,   compared to the micro-level thermal fluctuation in the level of  $10^{-20}$,  the  numerical  noises of  the three chaotic simulations  in the interval $[0,1000]$ are  much smaller and thus negligible.   Therefore, from the {\em mathematical} viewpoint, we are quite sure that  these  three  chaotic simulations are convergent to their {\em true}  trajectories.        

 \begin{figure}[t]
\centering
\includegraphics[scale=0.32]{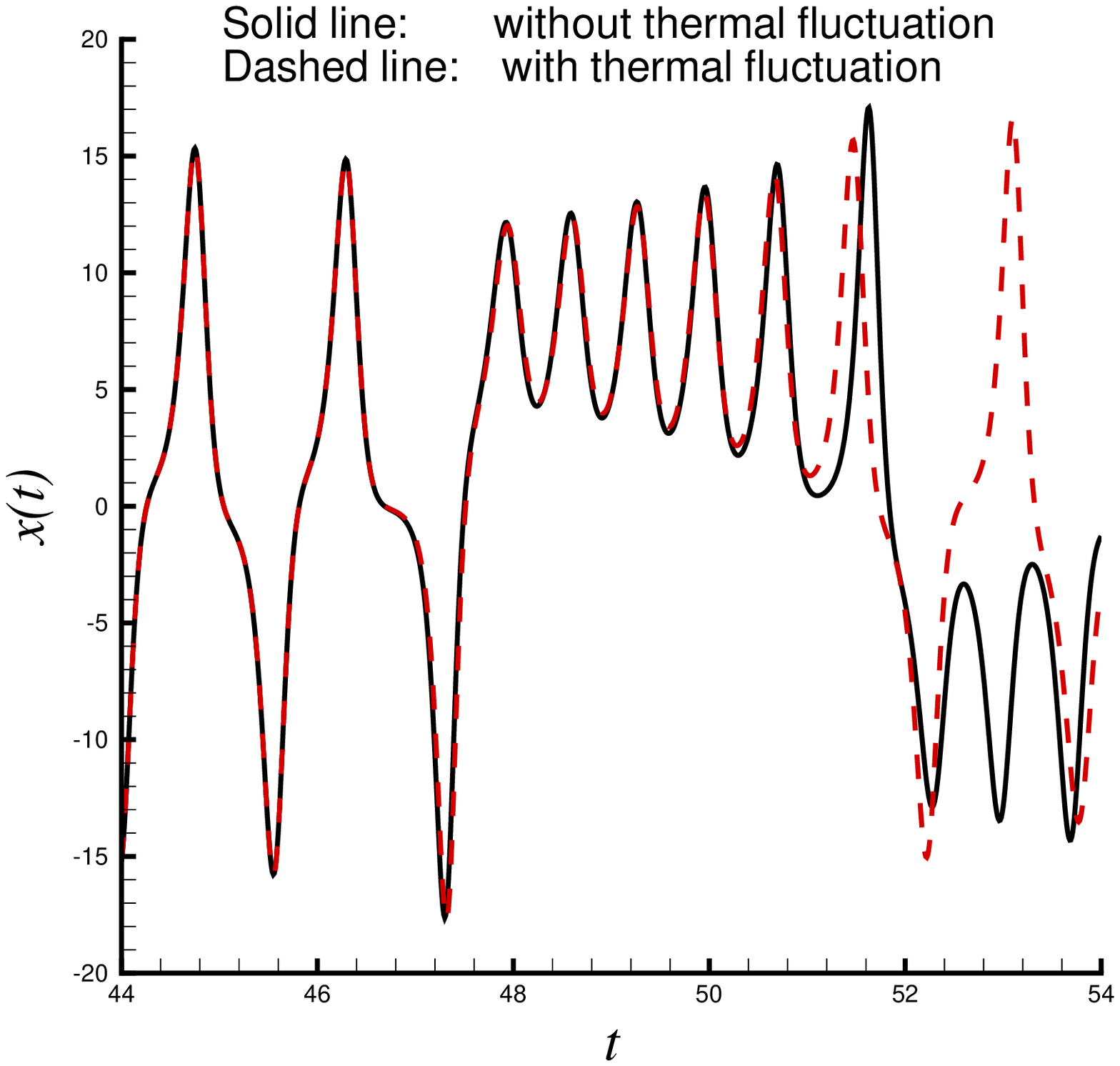}
\includegraphics[scale=0.32]{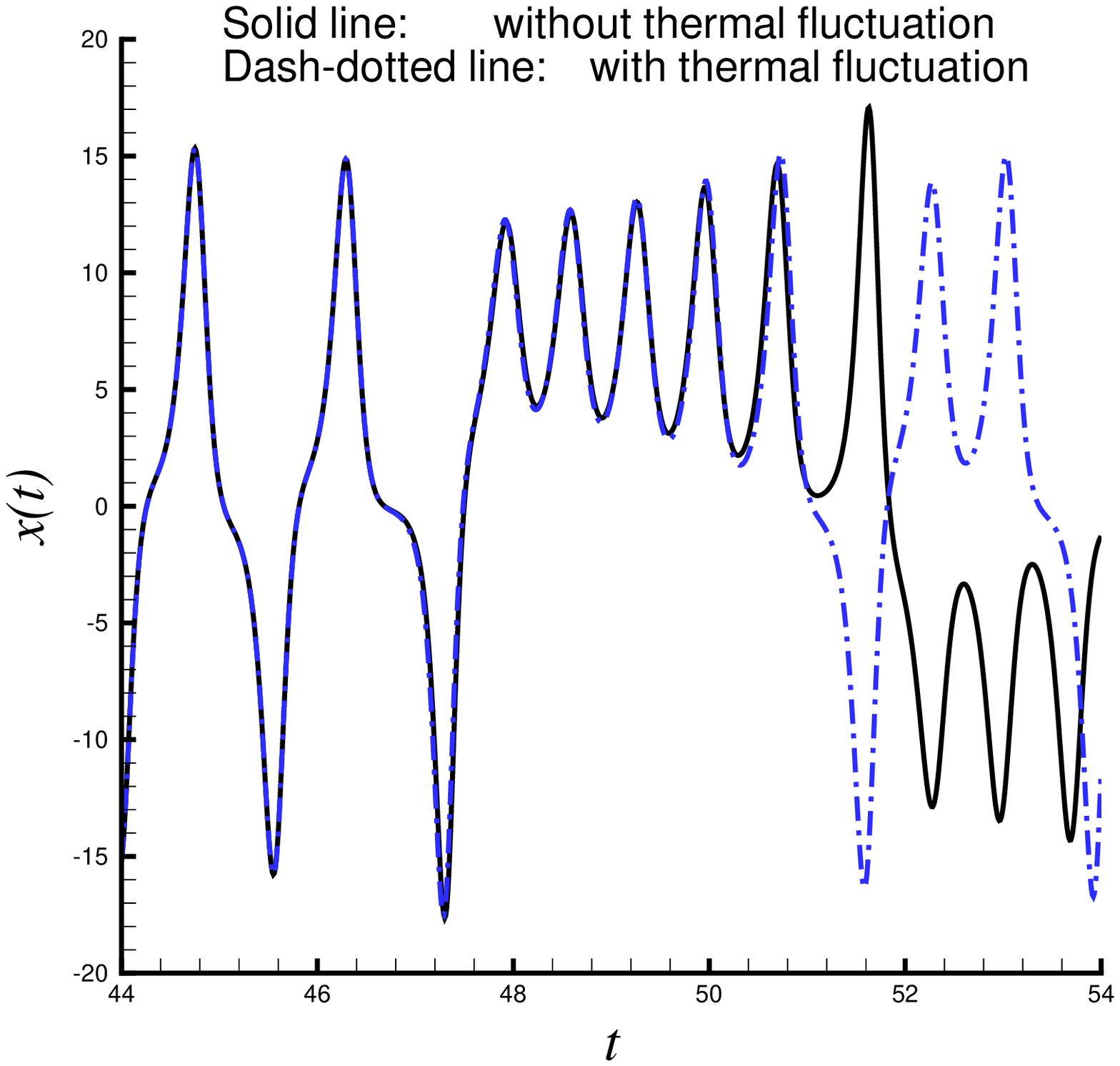}
\caption{Comparison of the reliable chaotic solutions by means of the CNS with $\Delta t = 10^{-2}$, the 400th-order Taylor expansion  and the 1000-digit multiple-precision (MP) data.  Solid line:   without thermal fluctuation using the initial condition (\ref{IC:A}); Dashed line:   with thermal fluctuation using the initial condition (\ref{IC:B}); Dash-dotted line:   with thermal fluctuation using the initial condition (\ref{IC:C}).     }
\label{figure:fluctuation}
\end{figure}       

However, as shown in Fig.~\ref{figure:fluctuation}, the three mathematically reliable chaotic simulations agree well only in the interval $t \in [0,51]$ but quickly depart from each other when $t > 51$.   Since, due to the thermal fluctuation mentioned above,  we do {\em not} know which initial condition among (\ref{IC:A}), (\ref{IC:B}) and (\ref{IC:C}) truly appears in nature,  all trajectories beyond $[0,51]$  are  uncertain.   Mathematically,  this numerical phenomenon is not surprising,  since  the ``butter-fly effect'' of chaos is well known.   However,  from the physical viewpoint,  
it means that  the random, inherent, micro-level thermal fluctuation has  a great  influence  on the  chaotic solutions!   It should be emphasized that thermal fluctuation is { inherently}  uncertain and {\em objective}, say,  it is independent of  any experimental accuracy of observation.   This suggests that,  although the reliable, accurate prediction of chaos in a finite but long enough interval is  {\em mathematically} possible,  the chaotic trajectories  for a long time might be essentially uncertain  in {\em physics}, mainly due  to  the the  objective,  unavoidable,  micro-level  thermal fluctuation.  

Therefore, due to the ``butterfly effect'' of chaotic dynamic systems like Lorenz equation,   the   inherent,  objective, physical,  micro-level uncertainty  (such as  the  thermal fluctuation) quickly truns into the macroscopic uncertainty, as revealed by our very accurate, reliable chaotic solutions mentioned above.    It  suggests that  trajectories  of chaotic  systems for a long time are essentially uncertain, from the physical viewpoint.  
Thus,  it might have no {\em physical}  meanings to give  a deterministic, accurate prediction of the  ``true''  trajectory of such a chaotic dynamic system, because  such  a kind  of  deterministic   trajectory does {\em not} exist at all in nature from the {\em physical} viewpoint.    Theoretically speaking,   such a kind  of  chaotic  dynamic systems  should  be  described  from the statistic viewpoints \cite{Liao2012CNSNS}.   Note that the same conclusions are obtained by Liao \cite{Liao2013CNSNS} for chaotic motions of three body problems,  who proposed a new concept ``physical limit of prediction''.   Thus, the above-mentioned conclusions  have the general meanings.

\section{Concluding remarks and discussions}

Using the parallel ``Clean Numerical Simulation'' (CNS) \cite{Liao2009, Liao2013} based on the 3500th-order of Taylor series method and data in the 4180-digit multiple precision,  we successfully obtain a mathematically reliable long-term prediction of chaotic solution of Lorenz equation in the interval $[0,10000]$  by means of  1200 CPUs of the National Supercomputer TH-1A at Tianjin, China.  To the best of our knowledge, this kind of  mathematically reliable chaotic solution in such a long interval has never been reported.  Mathematically, it provides us with a numerical benchmark for reliable long-term prediction of chaos.  Thus, given a finite but long enough interval, the CNS provides us with a safe tool to gain mathematically reliable chaotic solutions in it.  So, the CNS has important meanings not only in theory but also in practice. 

It is found that the  initial condition required for the mathematically reliable chaotic solution in [0,10000] must be in 4000-digit precision.  However, due to the inherent thermal fluctuation,  there exists the  physical uncertainty of the initial temperature at the level of $10^{-19}$, which is independent of  any experimental accuracy of observation or limited knowledge (in other words, it is the so-called objective uncertainty).  This objective, physical uncertainty is indeed rather small, but is much larger than the required 4000-digit precision of the initial temperature.   From the physical viewpoint, the three initial conditions  (\ref{IC:A}), (\ref{IC:B}) and (\ref{IC:C}) are the same,  since each of them may sample a state,   but  we do {\em not} know which ones truly appears in practice.   Due to the ``butter-fly effect'' of chaos, this  objective,  inherent,  micro-level uncertainty quickly turns into macroscopic uncertainty,  as shown in Fig.~\ref{figure:fluctuation}.    Note that such a  kind of relationship between micro-level and macroscopic uncertainty  is  supported  not only by  other physical models such as the chaotic motion of three-body problem  \cite{Liao2013CNSNS}  but also  some physical experiments  \cite{Bai1994, Xia2000}.     

Note that the Lorenz equation is derived from the Navier-Stokes equations with Boussinesq approximation in the macroscopic view, and thus the micro-level fluctuation of initial conditions should be negligible from the macroscopic viewpoint.  However, on the other side,  our reliable simulations of chaotic trajectories of Lorenz equation given by the CNS indicate that even the micro-level fluctuations of the initial conditions  can affect the  macroscopic  property  greatly.   This  leads   to  the so-called ``precision  paradox  of chaos'',   as  pointed  out  by  Liao \cite{Liao2009}, if  Lorenz equation itself is regarded  not only as  a  pure, mathematical model of chaos  but also a simplified, physical model related to Navier-Stokes equations.   In history, a paradox often implies something important and essential.   As  suggested by Liao \cite{Liao2013},  this paradox might imply the close relationship between the micro-level uncertainty and macroscopic randomness/uncertainty.    Obviously,  it  is  valuable  to   investigate  and  reveal   the  essence  of  this paradox in the future.    Note that,  there does not exist such kind of paradox for the chaotic motion of three-body problem,   but the same conclusions mentioned in this article   are obtained in \cite{Liao2013CNSNS}.

In summary,   it is {\em mathematically} indeed possible to gain convergent,  reliable  chaotic solutions  of Lorenz equation in a {\em finite} but long {\em enough} interval such as [0,10000].  However,  our very fine computations suggest that such kind of chaotic trajectories for a long time might be essentially uncertain in {\em physics},  due to  the  inherent, objective,   physical uncertainty of initial condition.

\section*{Acknowledgements}  

The first author would express his sincere acknowledgements to Prof. Yi-Long Bai (Institute of Mechanics, Chinese Academy of Sciences)  and Prof.   Meng-Fen Xia  (Department of Physics, Peking University), Prof. Hong-Ru Ma and Prof.  Li-Po  Wang (Shanghai Jiaotong University) for their  discussions on the micro-level and macroscopic uncertainty.   This work was carried out at National Supercomputer Center in Tianjin, China, and the calculations were performed on TH-1A.   This work is partly supported by National Natural Science Foundation of China under Grant No. 11272209, National Basic Research Program of China under Grant No. 2011CB309704, and State Key Laboratory of Ocean Engineering under Grant No. GKZD010056. 

\bibliography{nl_dynamics,viscous,homotopy,liao}

\bibliographystyle{unsrt}

\end{document}